\documentclass[12pt]{article}
\usepackage{graphicx}

\newcommand{\Pom}{I\!\!P}

\def\napoli{$^{a}$Institute of Nuclear Physics Polish Academy of Sciences,
Radzikowskiego 152, PL-31-342 Krak{\'o}w, Poland\\
$^{b}$Institut f\"ur Theoretische Physik, Universit\"at Heidelberg,
Philosophenweg 16, D-69120 Heidelberg, Germany}
\def\support{\footnote{Work supported by the MNiSW Grant No.~IP2014~025173 (Iuventus Plus) and the Polish National Science Centre grants DEC-2014/15/B/ST2/02528 and DEC-2015/17/D/ST2/03530.}}

\def\Title#1{\begin{center} {\Large #1 } \end{center}}
\def\Author#1{\begin{center}{ \sc #1} \end{center}}
\def\Address#1{\begin{center}{ \it #1} \end{center}}

\newcommand\pubblock{\rightline{\begin{tabular}{l} \\
         \end{tabular}}}
\newenvironment{Abstract}{\begin{quotation}  }{\end{quotation}}
\newenvironment{Presented}{\begin{quotation} \begin{center} 
             \end{center}\bigskip 
      \begin{center}\begin{large}}{\end{large}\end{center} \end{quotation}}

\def\beq{\begin{equation}}
\def\eeq#1{\label{#1}\end{equation}}
\def\eeqn{\end{equation}}

\def\beqa{\begin{eqnarray}}
\def\eeqa#1{\label{#1}\end{eqnarray}}
\def\eeqan{\end{eqnarray}}

\let\bar=\overbar

\def\Dslash{\not{\hbox{\kern-4pt $D$}}}
\def\dslash{\not{\hbox{\kern-2pt $\del$}}}

\def\msb{{\bar{\ssstyle M \kern -1pt S}}}

\begin{document}
\begin{titlepage}
\pubblock

\vfill
\Title{Exclusive diffractive production of hadrons in $pp$ collisions}
\vfill
\Author{Piotr Lebiedowicz$^{a}$\support, Otto Nachtmann$^{b}$, Antoni Szczurek$^{a}$}
\Address{\napoli}
\vfill
\begin{Abstract}
We discuss central exclusive diffractive dihadron production
in proton-proton collisions at high energies.
The calculation is based on a tensor pomeron model
and the amplitudes for the processes are formulated 
in an effective field-theoretic approach.
We include a purely diffractive dipion continuum, 
and the scalar and tensor resonances decaying into the $\pi^{+} \pi^{-}$ pairs
as well as the photoproduction contributions ($\rho^{0}$, Drell-S\"oding).
We discuss how two pomerons couple to tensor meson $f_{2}(1270)$
and the interference effects of the scalar and tensor resonances and the dipion continuum.
The theoretical results are compared with existing CDF and CMS experimental data.
We discuss also the Drell-Hiida-Deck type mechanism 
with centrally produced $\rho^{0}$ meson 
associated with a very forward/backward $\pi N$ system.
For the $pp \to pp \pi^{+} \pi^{-} \pi^{+} \pi^{-}$ reaction
we consider both the $\sigma \sigma$ and $\rho \rho$ contributions 
as well as the triple Regge exchange mechanism.
Predictions for planned or being carried out experiments (STAR, ALICE, ATLAS, CMS) are presented.
We show the influence of the experimental cuts on the integrated cross section
and on various differential distributions for outgoing particles.



\end{Abstract}
\vfill
\begin{Presented}
Presented at EDS Blois 2017, Prague, \\ Czech Republic, June 26-30, 2017
\end{Presented}
\vfill
\end{titlepage}
\def\thefootnote{\fnsymbol{footnote}}
\setcounter{footnote}{0}

\section{Introduction}
Central production mediated by the ``fusion'' of two exchanged pomerons
is an important diffractive process for the investigation of properties of resonances,
in particular, for search of gluonic bound states (glueballs) \cite{Lebiedowicz:2016ioh,Fiore:2015lnz}.
This is closely related to experimental studies of the COMPASS \cite{Austregesilo:2016sss}, 
ISR \cite{Breakstone:1986xd},
STAR \cite{Adamczyk:2014ofa}, 
CDF \cite{Aaltonen:2015uva}, 
LHCb \cite{McNulty:2016sor},
ALICE \cite{Schicker:2012nn},
CMS \cite{Khachatryan:2017xsi},
ATLAS \cite{Staszewski:2011bg}.
On the theoretical side, the exclusive diffractive dihadron continuum production 
can be understood as due to the exchange of two pomerons
between the external protons and the centrally produced hadronic system.
One of the first calculations were concerned with the $p p \to p p \pi^+ \pi^-$ reaction 
\cite{Lebiedowicz:2009pj,Lebiedowicz:2011nb,Lebiedowicz:2015eka}.
The Born amplitude 
was written in terms of pomeron/reggeon exchanges 
with parameters fixed from phenomenological analyses of
$NN$ and $\pi N$ total and elastic scattering.
Such calculations make sense for the continuum production
of pseudoscalar meson pairs.
First calculations of central exclusive diffractive production of $\pi^{+} \pi^{-}$ continuum
together with the dominant scalar $f_{0}(500)$, $f_{0}(980)$, 
and tensor $f_{2}(1270)$ resonances 
were performed in \cite{Lebiedowicz:2016ioh} within the tensor-pomeron model formulated in \cite{Ewerz:2013kda}.
In this model the soft pomeron exchange 
is described as an effective rank-2 symmetric tensor exchange; 
see \cite{Nachtmann:1991ua,Ewerz:2013kda}.
There, all reggeon exchanges with charge conjugation $C = +1$ ($C = -1$) were described
as effective tensor (vector) exchanges.
Recently, in \cite{Ewerz:2016onn}, three models for the soft pomeron, tensor, vector, and scalar, 
were compared with the STAR data on polarised high-energy $pp$ scattering \cite{Adamczyk:2012kn}.
Only the tensor-pomeron model was found to be consistent with
the general rules of quantum field theory and the data from \cite{Adamczyk:2012kn}.
This model was applied to the diffractive production of several scalar and pseudoscalar mesons 
in the reaction $p p \to p p M$; see \cite{Lebiedowicz:2013ika}.

The exclusive $\rho^0$ and non-resonant (Drell-S\"oding)
photon-pomeron/reggeon $\pi^{+} \pi^{-}$ production in $pp$ collisions 
was studied in \cite{Lebiedowicz:2014bea}.
In \cite{Lebiedowicz:2014bea} we showed that the resonant contribution
interferes with the non-resonant $\pi^{+} \pi^{-}$ continuum
leading to a skewing of the $\rho(770)$-meson line shape; see also \cite{Bolz:2014mya}.
Due to the photon propagators occurring in diagrams,
we expect these processes to be most important when at least one of the protons
undergoes only a very small $|t_{1,2}|$.
We discussed also the Drell-Hiida-Deck type mechanism with centrally produced
$\rho^{0}$ meson associated with a very forward/backward $\pi N$ system
in the $pp \to pp \rho^{0} \pi^{0}$ and $pp \to pn \rho^{0} \pi^{+}$ reactions \cite{Lebiedowicz:2016ryp}.


\section{Sketch of formalism}

The total diffractive amplitude is a sum of continuum amplitude 
and the amplitudes with the $s$-channel resonances (with quantum numbers $I^{G}J^{PC}=0^{+}({\rm even})^{++}$):
%
\begin{eqnarray} \label{amplitude_pomTpomT}
&&{\cal M}_{pp \to pp \pi^{+} \pi^{-}} =
{\cal M}^{\pi \pi{\rm-continuum}}_{pp \to pp \pi^{+} \pi^{-}} + 
{\cal M}^{(\Pom \Pom \to f_{0} \to \pi^{+}\pi^{-})}_{pp \to pp \pi^{+}\pi^{-}}
 + {\cal M}^{(\Pom \Pom \to f_{2} \to \pi^{+}\pi^{-})}_{pp \to pp \pi^{+}\pi^{-}}\,.
\end{eqnarray}
For instance, the Born (no absorption effects) amplitude for the process
$pp \to pp (\Pom \Pom \to f_{2} \to \pi^{+}\pi^{-})$ can be written as
\begin{eqnarray}
&&{\cal M}^{(\Pom \Pom \to f_{2}\to \pi^{+}\pi^{-})}_{\lambda_{a} \lambda_{b} \to \lambda_{1} \lambda_{2} \pi^{+}\pi^{-}} 
=  (-i)\,
\bar{u}(p_{1}, \lambda_{1}) 
i\Gamma^{(\Pom pp)}_{\mu_{1} \nu_{1}}(p_{1},p_{a}) 
u(p_{a}, \lambda_{a})\;
i\Delta^{(\Pom)\, \mu_{1} \nu_{1}, \alpha_{1} \beta_{1}}(s_{1},t_{1}) \nonumber \\
&& \qquad \qquad \qquad \qquad \times 
i\Gamma^{(\Pom \Pom f_{2})}_{\alpha_{1} \beta_{1},\alpha_{2} \beta_{2}, \rho \sigma}(q_{1},q_{2}) \;
i\Delta^{(f_{2})\,\rho \sigma, \alpha \beta}(p_{34})\;
i\Gamma^{(f_{2} \pi \pi)}_{\alpha \beta}(p_{3},p_{4}) \nonumber \\
&& \qquad \qquad \qquad \qquad \times 
i\Delta^{(\Pom)\, \alpha_{2} \beta_{2}, \mu_{2} \nu_{2}}(s_{2},t_{2}) \;
\bar{u}(p_{2}, \lambda_{2}) 
i\Gamma^{(\Pom pp)}_{\mu_{2} \nu_{2}}(p_{2},p_{b}) 
u(p_{b}, \lambda_{b}) \,,
\label{amplitude_f2_pomTpomT}
\end{eqnarray}
where $t_{1} =(p_{1} - p_{a})^{2}$, $t_{2} =(p_{2} - p_{b})^{2}$, $s_{1} = (p_{a} + q_{2})^{2} = (p_{1} + p_{34})^{2}$,
$s_{2} = (p_{b} + q_{1})^{2} = (p_{2} + p_{34})^{2}$,
$p_{34} = p_{3} + p_{4}$. $\Delta^{(\Pom)}$ and $\Gamma^{(\Pom pp)}$ 
denote the effective propagator and proton vertex function, respectively, 
for the tensor pomeron exchange.
For the explicit expressions see section~3 of \cite{Ewerz:2013kda}.
In \cite{Lebiedowicz:2016ioh} we considered all (seven) possible tensorial structures 
for the $\Pom \Pom f_{2}$ coupling.
Other details as form of form factors, the tensor-meson propagator $\Delta^{(f_{2})}$ and 
the $f_{2} \pi \pi$ vertex are given in \cite{Ewerz:2013kda,Lebiedowicz:2016ioh}.
In reality the Born approximation is usually not sufficient and 
absorption corrections (rescattering effects) have to be taken into account,
see e.g. \cite{Harland-Lang:2013dia,Lebiedowicz:2015eka}.
The absorption effects due to $pp$ and $\pi p$ interactions, 
discussed in \cite{Lebiedowicz:2015eka},
lead to a significant modification of the shape of the distributions 
in $\phi_{pp}$, $p_{t,p}$, $t_{1,2}$ and could be tested by the 
CMS-TOTEM and ATLAS-ALFA groups.

\section{Some selected results}

\begin{figure}[htb]
 \centering
(a)  \includegraphics[height=2.5in]{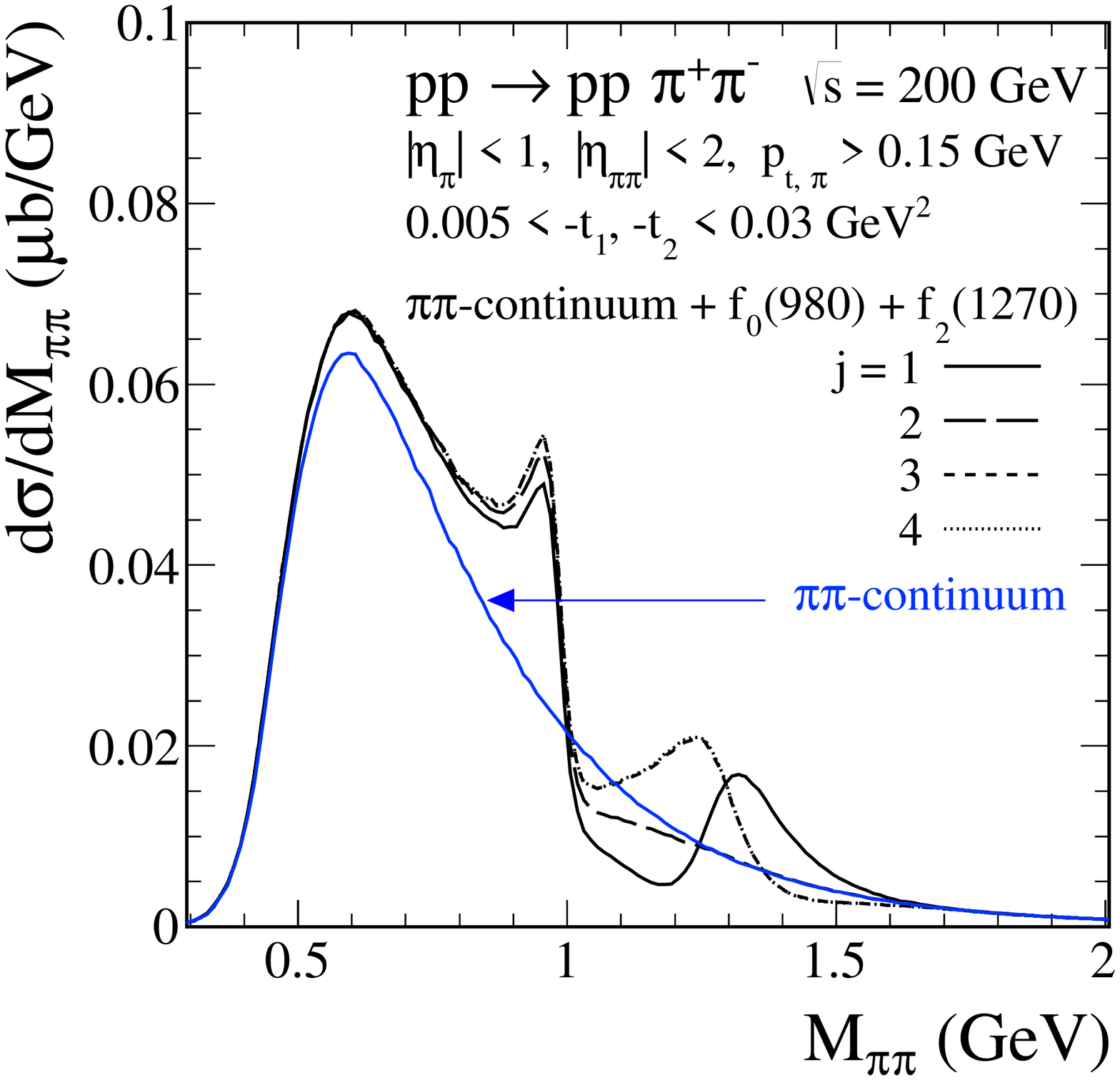}
(b)  \includegraphics[height=2.5in]{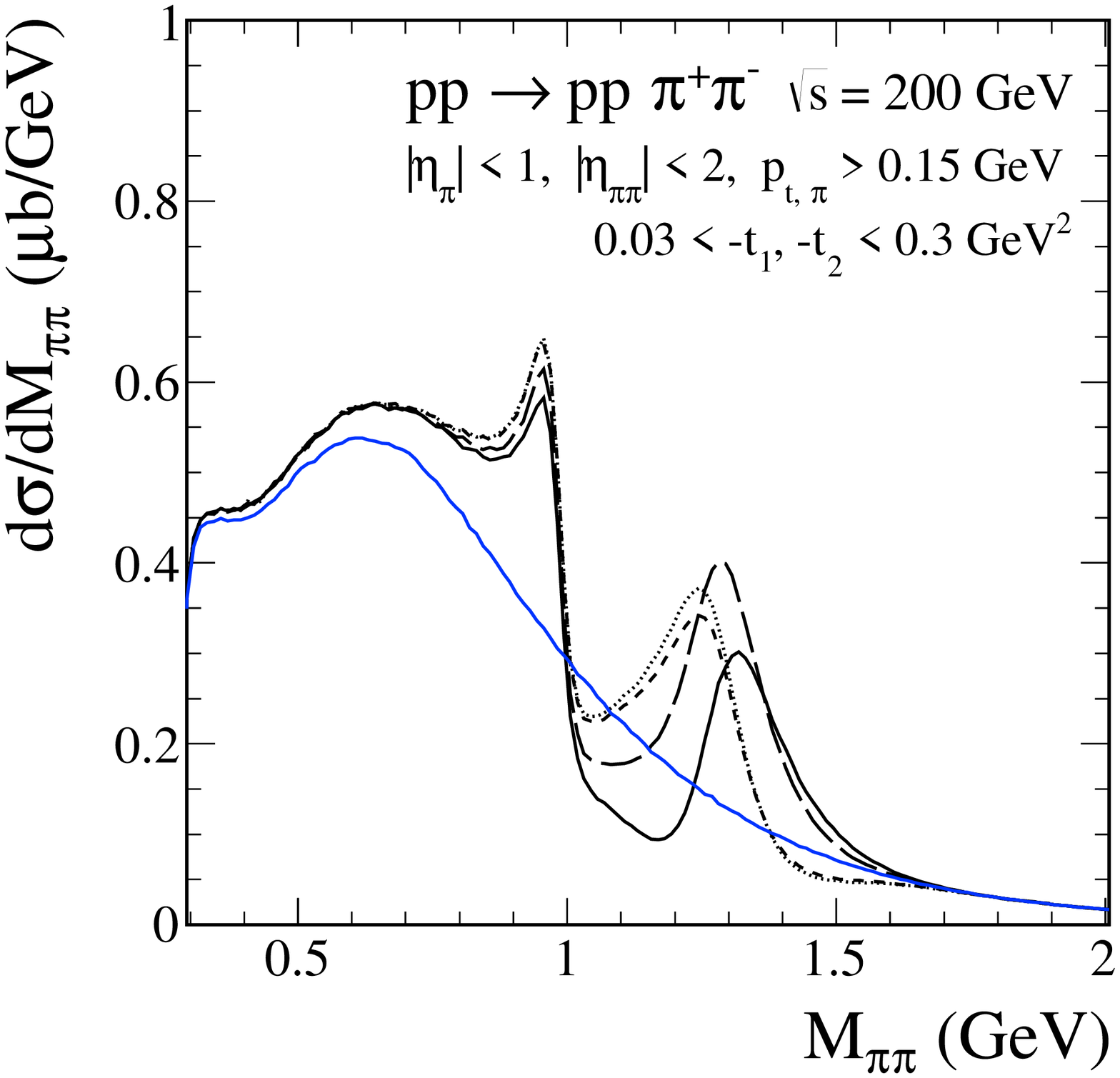}
(c)  \includegraphics[height=2.5in]{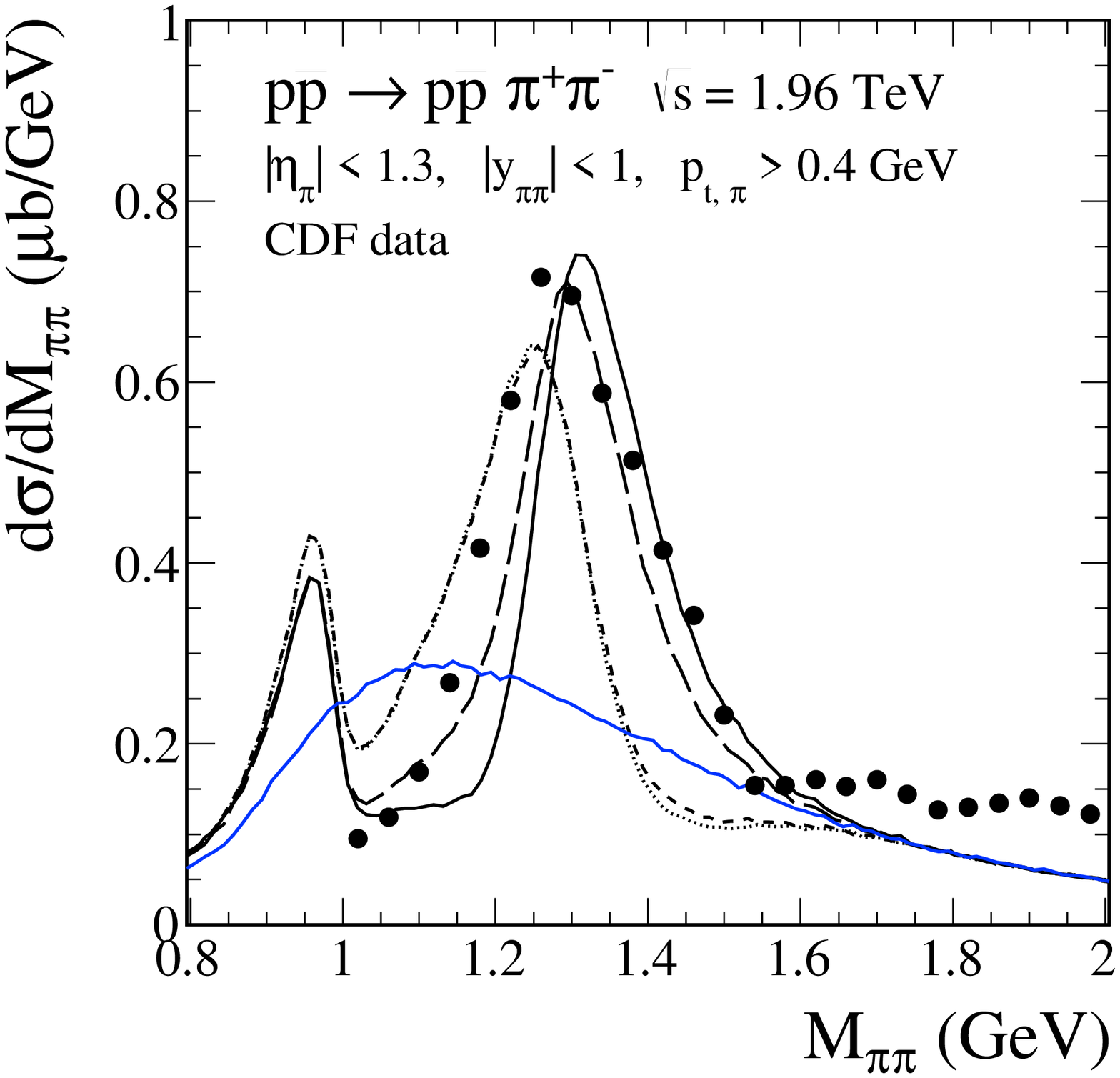}
(d)  \includegraphics[height=2.5in]{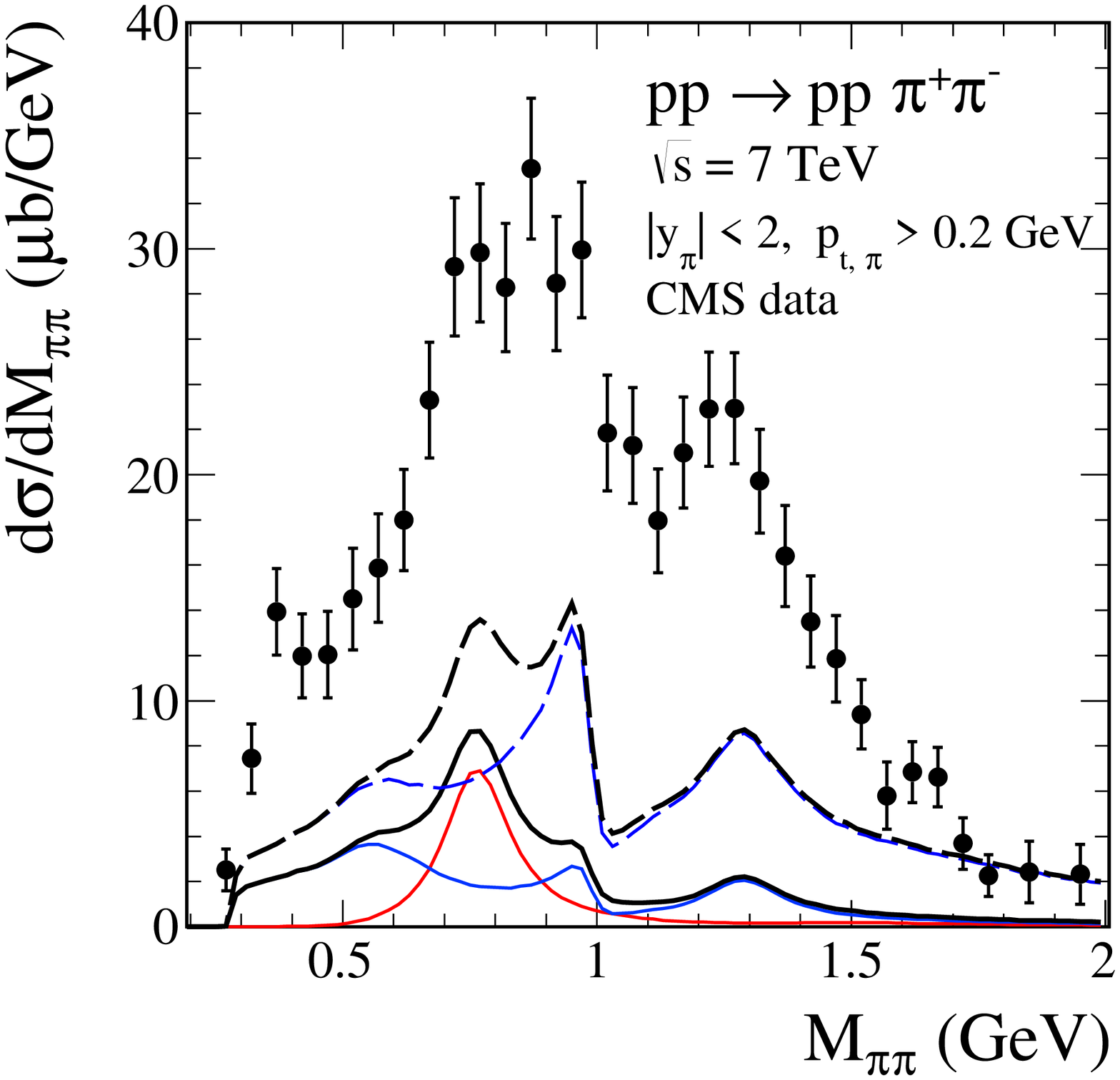}
  \caption{Two-pion invariant mass distributions 
for the STAR \cite{Adamczyk:2014ofa} ((a) and (b) panels),
the CDF \cite{Aaltonen:2015uva} ((c) panel)
and the CMS \cite{Khachatryan:2017xsi} ((d) panel) kinematics.
In the panels (a) - (c) the individual contributions 
of different $\Pom \Pom f_{2}$ couplings ($j = 1, ..., 4$) are shown.
The Born calculations for $\sqrt{s}=0.2, 1.96$~TeV were multiplied 
by the gap survival factor $\langle S^{2}\rangle = 0.2, 0.1$, respectively.
The blue solid lines represent the non-resonant continuum contribution only 
($\Lambda_{off,M} = 0.7$~GeV) 
while the black lines represent a coherent sum of non-resonant continuum, 
$f_{0}(980)$ and $f_{2}(1270)$ resonant terms.
The CDF data are from \cite{Aaltonen:2015uva}.
In the panel (d) both the photoproduction (red line) 
and purely diffractive (blue line) contributions 
multiplied by the factors $\langle S^{2}\rangle = 0.9$
and $\langle S^{2}\rangle = 0.1$, respectively, are included.
The complete results correspond to the black solid line ($\Lambda_{off,M} = 0.7$~GeV)
and the black dashed line ($\Lambda_{off,M} = 1.2$~GeV).
The CMS preliminary data scanned from \cite{Khachatryan:2017xsi} are shown for comparison.
}
\label{fig:fig1}
\end{figure}

In \cite{Lebiedowicz:2016ioh} we tried to describe
the dipion invariant mass distribution observed by different experimental groups.
As can be clearly seen from Fig.~\ref{fig:fig1}~(a) - (c)
different $\Pom \Pom f_{2}$ couplings generate different interference patterns
around $M_{\pi\pi} \sim 1.27$~GeV.
A sharp drop around $M_{\pi\pi} \sim 1$~GeV is attributed to the interference
of $f_{0}(980)$ and continuum.
We found that the shape of the $\pi^{+}\pi^{-}$ distribution depends
on the cuts used in a particular experiment
(usually the $t$ cuts are very different for different experiments).
We can observe that the $j=2$ coupling gives results close to those observed by the CDF Collaboration \cite{Aaltonen:2015uva}.
In this preliminary study we did not try to fit the CDF data \cite{Aaltonen:2015uva} 
by mixing different couplings because the data are not fully exclusive
(the outgoing $p$ and $\bar{p}$ were not measured).
The calculations were done at Born level
and the absorption corrections were taken into account
by multiplying the cross section
by a common factor $\langle S^{2}\rangle$ obtained from \cite{Lebiedowicz:2015eka}.
The two-pion continuum was fixed by choosing a form factor for the off-shell pion 
$\hat{F}_{\pi}(k^{2})=\frac{\Lambda^{2}_{off,M} - m_{\pi}^{2}}{\Lambda^{2}_{off,M} - k^{2}}$ and
$\Lambda_{off,M} = 0.7$~GeV.

In panel~(d) of Fig.~\ref{fig:fig1} we show results including 
in addition to the non-resonant $\pi^{+}\pi^{-}$ continuum, 
the $f_{2}(1270)$ and the $f_{0}(980)$ resonances,
the contribution from photoproduction ($\rho^{0} \to \pi^{+}\pi^{-}$, Drell-S\"oding mechanism), 
as well as the $f_{0}(500)$ resonant contribution.
Our predictions are compared with the CMS preliminary data \cite{Khachatryan:2017xsi}.
We assume only one of the seven possible $\Pom \Pom f_{2}$ tensorial couplings, 
that is $j =2$ coupling;
see \cite{Lebiedowicz:2016ioh}. 
Here the absorption effects lead to a huge damping of the cross section 
for the purely diffractive term (the blue lines) and relatively small
reduction of the cross section for the photoproduction term (the red lines).
Therefore we expect one could observe the photoproduction contribution.
The CMS measurement \cite{Khachatryan:2017xsi} is not fully exclusive and
the experimental spectra contain contributions associated
to other processes, e.g., when one or both protons undergo dissociation.
In addition we show that the results with $\Lambda_{off,M} = 1.2$~GeV
better describe the preliminary CMS data (see the dashed line).
If we used this set for the STAR or CDF measurements
our results there would be above the preliminary STAR data \cite{Adamczyk:2014ofa}
at $M_{\pi\pi} > 1$~GeV and in complete disagreement 
with the CDF data from \cite{Aaltonen:2015uva}.
Only purely central exclusive data expected from
CMS-TOTEM and ATLAS-ALFA will allow to draw definite conclusions.

\section{Summary of our recent results}

In \cite{Lebiedowicz:2016ioh} 
we have analysed the exclusive central production of
dipion continuum and resonances contributing to the $\pi^{+} \pi^{-}$ pair production
in proton-(anti)proton collisions in an effective field-theoretic approach
with tensor pomerons and reggeons proposed in \cite{Ewerz:2013kda}.
We have included the scalar $f_{0}(500)$ and $f_{0}(980)$ resonances,
the tensor $f_{2}(1270)$ resonance and the vector $\rho(770)$ resonance in a consistent way.
In the case of $f_{2}(1270)$-meson production via ``fusion'' of two tensor pomerons
we have found \cite{Lebiedowicz:2016ioh} all possible $\Pom \Pom f_{2}$ tensorial couplings.
The different couplings give different results due to 
different interference effects of the $f_{2}$ resonance 
and the dipion continuum contributions. 
We have shown that the resonance structures in the measured two-pion invariant mass spectra
depend on the cut on proton transverse momenta and/or 
on four-momentum transfer squared $t_{1,2}$ used in an experiment.
The cuts may play then the role of a $\pi \pi$ resonance filter.
The model parameters of the optimal $\Pom \Pom f_{2}$ coupling ($j=2$) have been roughly adjusted
to the recent CDF and preliminary STAR experimental data 
and then used for the predictions for the ALICE, and CMS experiments.
We have made estimates of cross sections for both the diffractive 
and photoproduction contributions and 
we have presented several interesting correlation distributions
which could be checked by the experiments.

The $pp \to pN \rho^{0} \pi$ processes
constitute an inelastic (non-exclusive) background 
to the $p p \to p p \rho^0$ reaction in the case when 
only the centrally produced $\rho^{0}$ meson 
decaying into $\pi^{+} \pi^{-}$ is measured,
the final state protons are not observed, and only 
rapidity-gap conditions are checked experimentally.
We have estimated the size of the proton diffractive-dissociative background
to the exclusive $p p \to p p \rho^0$ process at LHC energies \cite{Lebiedowicz:2016ryp}.
The ratio of integrated cross sections 
for the inelastic $p p \to p N \rho^0 \pi$ processes
to the reference reaction $pp \to \ pp \rho^0$ is of order of (7--10)\%.


Also the studies of different decay channels in central
exclusive production would be very valuable.
One of the possibilities is the reaction $p p \to p p \pi^+ \pi^- \pi^+ \pi^-$
being analysed at the LHC.
In Ref.~\cite{Lebiedowicz:2016zka} 
we analysed the reaction $pp \to pp \pi^{+} \pi^{-} \pi^{+} \pi^{-}$
proceeding via the intermediate $\sigma \sigma$ and the $\rho \rho$ states.
The results for processes with the exchange of heavy mesons (compared to pion)
strongly depend on the details of the hadronic form factors.
By comparing the theoretical results and the cross sections
found in the ISR experiment \cite{Breakstone:1993ku}
we fixed the parameters of the off-shell meson form factor
and the $I\!\!P \sigma \sigma$ and $f_{2 I\!\!R} \sigma \sigma$ couplings.
The exclusive $\pi^{+}\pi^{-}\pi^{+}\pi^{-}$ 
continuum production was discussed recently in \cite{Kycia:2017iij}.
A measurable cross section of order of a few $\mu b$ was obtained
including the experimental cuts relevant for the LHC experiments.

\end{document}